\documentclass[preprint,preprintnumbers,amsmath,amssymb,superscriptaddress]{revtex4}

    \usepackage{amsmath}
    \usepackage{makeidx}
    \usepackage{amsfonts}
    \usepackage[ansinew]{inputenc}
    \usepackage{subfigure}
    \usepackage{epsfig}
    \usepackage[colorlinks,hyperindex]{hyperref}
    \hypersetup
    {
        colorlinks,%
        citecolor=black,%
        linkcolor=black,%
        urlcolor=black,%
    }

\usepackage{color}
\usepackage{graphicx}
\usepackage{dcolumn}
\usepackage{amsmath}    
\usepackage{verbatim}   
\usepackage{color}      
\usepackage{hyperref}   
\usepackage{amsfonts}
\usepackage{braket}
\usepackage{bm}
\usepackage{epstopdf}
\usepackage{float}

\begin{document}


\title{Ramsey Interference in a Multi-level Quantum System}

\author{J. P. Lee}\,
\affiliation{Toshiba Research Europe Limited, Cambridge Research Laboratory,\\
208 Science Park, Milton Road, Cambridge, CB4 0GZ, U. K.}
\affiliation{Engineering Department, University of Cambridge,\\
9 J. J. Thomson Avenue, Cambridge, CB3 0FA, U. K.}

\author{A. J. Bennett}
\email{anthony.bennett@crl.toshiba.co.uk}
\affiliation{Toshiba Research Europe Limited, Cambridge Research Laboratory,\\
208 Science Park, Milton Road, Cambridge, CB4 0GZ, U.
K.}

\author{J. Skiba-Szymanska}
\affiliation{Toshiba Research Europe Limited, Cambridge Research Laboratory,\\
208 Science Park, Milton Road, Cambridge, CB4 0GZ, U. K.}

\author{D. J. P. Ellis}
\affiliation{Toshiba Research Europe Limited, Cambridge Research Laboratory,\\
208 Science Park, Milton Road, Cambridge, CB4 0GZ, U. K.}

\author{I. Farrer}
\affiliation{Cavendish Laboratory, Cambridge University,\\
J. J. Thomson Avenue, Cambridge, CB3 0HE, U. K.}

\author{D. A. Ritchie}
\affiliation{Cavendish Laboratory, Cambridge University,\\
J. J. Thomson Avenue, Cambridge, CB3 0HE, U. K.}

\author{A. J. Shields}
\affiliation{Toshiba Research Europe Limited, Cambridge Research Laboratory,\\
208 Science Park, Milton Road, Cambridge, CB4 0GZ, U. K.}

\date{\today}%


\begin{abstract}
We report Ramsey interference in the excitonic population of a
negatively charged quantum dot revealing the coherence of the state
in the limit where radiative decay is dominant. Our experiments show
that the decay time of the Ramsey interference is limited by the
spectral width of the transition. Applying a vertical magnetic field
induces Zeeman split transitions that can be addressed by changing
the laser detuning to reveal 2, 3 and 4 level system behaviour. We
show that under finite field the phase-sensitive control of two
optical pulses from a single laser can be used to prepare both
population and spin qubits simultaneously.
\end{abstract}

\maketitle 




Ramsey interferometry has found use in caesium atomic clocks
and in investigations of the quantum nature of the electromagnetic
field \cite{essen1955atomic,haroche2006exploring}. Using this high precision technique to study zero dimensional defects in solid
state systems promises new applications and insights
\cite{ramsay2010review,jayakumar2013deterministic,litvinenko2015coherent,de2010universal}.
Prior measurements of RI between exciton population levels in
quantum dots (QDs) have used one of two techniques. In some works
\cite{bonadeo1998coherent,stufler2006ramsey, ramsay2010review}, the
laser is resonant with a transition and measurement of the population
is made in photocurrent, which necessitates the destruction of the
excition, shortening its lifetime. Alternatively, optical readout of
the population has been made when excitation occurs via a
phonon-assisted transition \cite{bonadeo1998coherent} or a
two-photon transition \cite{jayakumar2013deterministic} and thus the
laser can be removed by spectral filtering.

\begin{figure}[h]
\includegraphics[width=70mm]{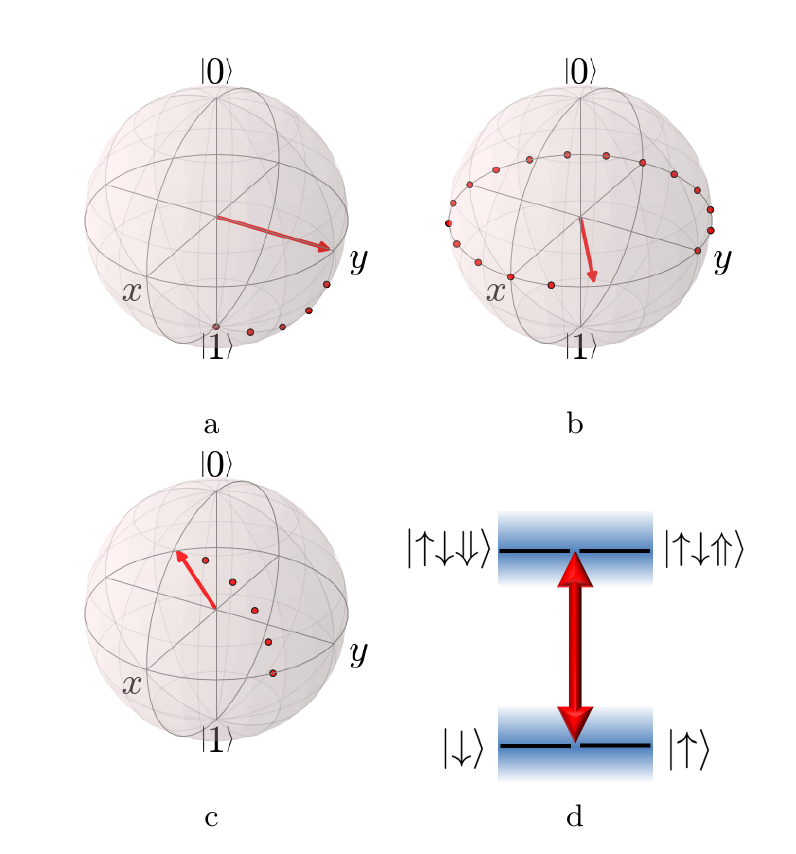}
\caption{\label{Fig1} {\bf a)} A $\frac{\pi}{2}$ pulse creates a
coherent superposition of the excited and ground states; the Bloch
vector is rotated by $\frac{\pi}{2}$ about the x-axis. {\bf b)} The
system undergoes free evolution; the Bloch vector precesses about
the z-axis. {\bf c)} A second $\frac{\pi}{2}$ pulse maps the system
into the excited or ground state dependent on the phase difference
between the pulses; the Bloch vector is rotated by $\frac{\pi}{2}$
about the x-axis. {\bf d)} In zero magnetic field the negatively
charged QD behaves as a two level system. (Bloch spheres were
produced with \cite{johansson2013qutip}). }
\end{figure}

Here we demonstrate a direct measurement of RI in the resonant
fluorescence (RF) of a QD, building on the recent demonstration of
pulsed RF \cite{he2013demand}. The measurements are performed on the
negative trion transition and dependent on the external magnetic
field and laser energy, we observe 2, 3 and 4 level system
behaviour. This leads to a rich variety interferograms with
exponential decays, beats and oscillations which reveal information
about the coherence of the exciton and the trapped charge. Our
results show how a single laser can be used to prepare a population
qubit simultaneously with a spin qubit, opening new possibilities to
control and measure the dynamics of spin-qubits in mesoscopic
semiconductor systems.

Ramsey interference (RI) fringes are the result of the oscillation in the
population of a state of as a function of the phase difference between
two $\frac{\pi}{2}$ pulses. The first pulse creates a
superposition between the ground and excited state -
interpreted as a $\frac{\pi}{2}$ rotation of the Bloch vector about
the x-axis of the Bloch sphere (Figure \ref{Fig1}a). The system then
undergoes free evolution - the Bloch vector precesses about
the equator of the Bloch sphere (Figure \ref{Fig1}b) at a frequency
determined by the energy difference between the ground and excited
states. The second $\frac{\pi}{2}$ pulse then maps then the system
to the ground or excited state dependent on the phase difference of
the two pulses. Again, this is interpreted as a rotation of the
Bloch vector about the x-axis (Figure \ref{Fig1}c). At zero magnetic field the
energy levels of a negatively charged QD
are spin independent, so the dot behaves as a
two level system (Figure \ref{Fig1}d).

\begin{figure}[H]
\includegraphics[width=160mm]{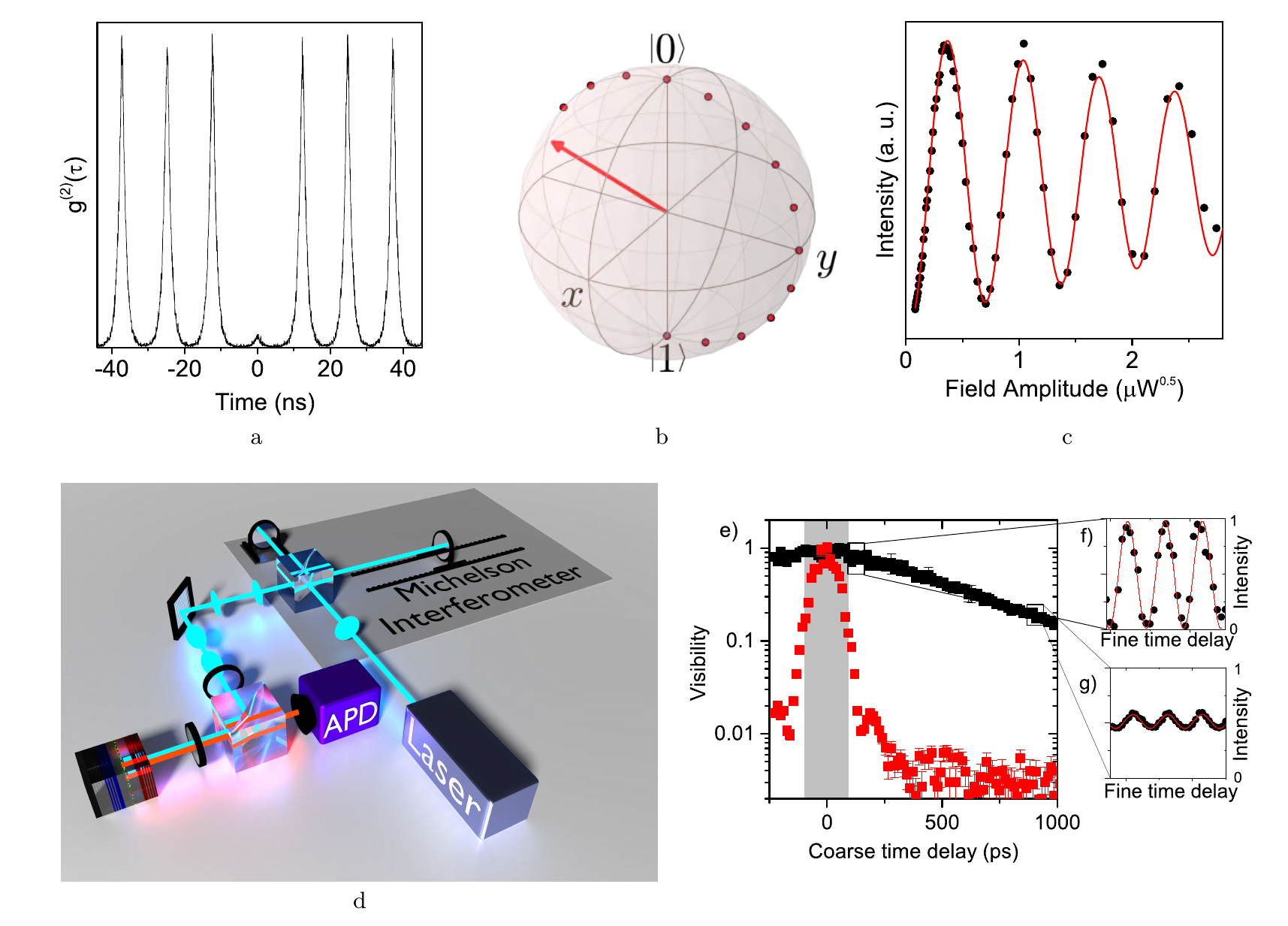}
\caption{\label{Fig2} {\bf a)} A pulsed Hanbury-Brown and Twiss
$g^{(2)}(\tau)$ measurement using $0.7\pi$-pulses. The absence of a
peak at $\tau=0$ is indicative of a single photon source. {\bf b)}
The rotation of the Bloch vector about the x-axis. {\bf c)} The
emitted light intensity as a function of the square root of pulse
power shows several Rabi oscillations. {\bf d)} An illustration of
our experimental setup for performing RI. Two pulses with a variable
separation are created by a Michelson interferometer and directed to
the QD-Light Emitting Diode (QD-LED) via a Polarising Beam Splitter
(PBS) and a $\frac{\lambda}{4}$-waveplate. Light from the QD-LED is
polarisation filtered by the PBS and collected by an avalanche
photo-diode. {\bf e)} (Red) Power measured in front of the sample.
(Black) The visibility of the Ramsey fringes as a a function pulse
separation. The grey area of the graph indicates the range of pulse
separations over which there is a significant overlap between the
input laser pulses. {\bf f)} A sinusoidal oscillation in emission
intensity as a function of fine time delay at a short pulse
separation. {\bf g)} A sinusoidal oscillation in the emission
intensity as a function of fine time delay at a longer pulse
separation.}
\end{figure}

The InGaAs dot used was embedded in the undoped
region of an LED planar microcavity cooled to 5K. The device was
held below the the threshold voltage so no electroluminescence was
seen, but the dot was loaded with an electron. Excitation was
performed by pulsed RF with 57 ps pulses,
which are short relative to the exciton lifetime (955$\pm$1 ps).
Cross-polarisation filtering was used to separate the excitation
light from the emitted light and the emitted light was not
spectrally filtered. This ensured that the spectral density function
of the emitted photons remained intact, preventing an artificial
increase in the decay time of the Ramsey fringe visibility. To
determine that we are exciting a single quantum emitter we perform a
Hanbury-Brown and Twiss measurement and record $g^{(2)}(0)=0.06$
(Figure \ref{Fig2}a).

Next, we demonstrate that we can coherently manipulate the
excitation population by rotating the Bloch vector about the x-axis
(Figure \ref{Fig2}b). We do this by recording the emitted light
intensity as a function of the square root of pulse power and
observing Rabi oscillations with some damping
\cite{melet2008resonant,flagg2009resonantly,ramsay2010damping}
(Figure \ref{Fig2}c).

We use the setup illustrated in Figure \ref{Fig2}d to apply the
Ramsey pulse sequence and record the emitted light intensity as a
function of pulse separation. Scanning the position of the moveable
mirror of the Michelson interferometer allows us to control the
pulse separation and to record the emitted light intensity after
polarisation filtering using a silicon avalanche-photodiode. We
observe a sinusoidal variation in the emission intensity (Figures
\ref{Fig2}f and \ref{Fig2}g) from which we can extract the
visibility at each pulse separation time (\ref{Fig2}e in black). The
visibility decays exponentially with a decay time of 510$\pm$10 ps.
The decay time is close to the coherence time inferred from a
linewidth measurement (600$\pm$20 ps), as expected from the
Wiener-Khinchin Theorem (\cite{skantzakis2010tracking}). Figure
\ref{Fig2}e also shows the interference between the laser pulses
measured in front of the sample (red). Due to this effect,
oscillations shown in the grey region cannot be ascribed to RI.
Outside of this region the low noise floor indicates that the
oscillations are a result of RI.

\begin{figure}[h]
\includegraphics[width=80mm]{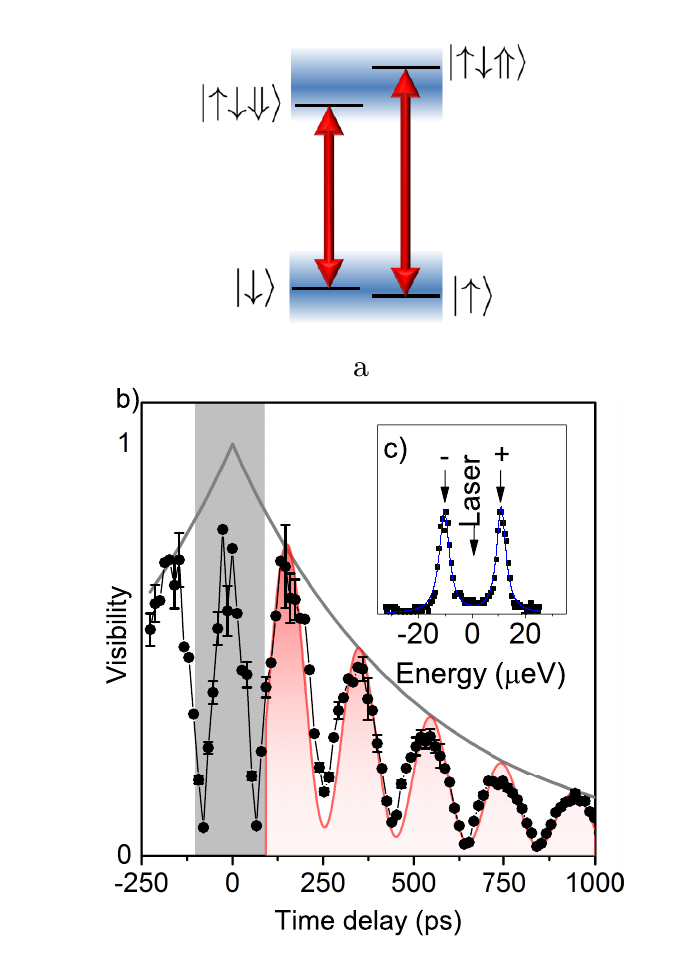}
\caption{\label{Fig3} {\bf a)} In a Faraday magnetic field there are
two allowed transitions. {\bf b)} A beating in the Ramsey fringe
visibility at 125 mT corresponding to the energy difference between
the two allowed transitions is seen inside an exponentially decaying
envelope. The grey area indicates the region of overlap between the
input pulses and the grey exponential decay is a fit to the decay
envelope of the 2-level system as in Figure \ref{Fig2}e. {\bf c)}
The energy of the excitation laser is tuned to excite both allowed
transitions. }
\end{figure}

Applying a magnetic field in the Faraday geometry removes the
degeneracy between the spin-up and spin-down states resulting in
4-levels with distinct energies (Figure \ref{Fig3}a). There are now
two transitions allowed by the selection rules with an energy
difference consistent with a combined electron and hole g-factor of
$|g_{e}+g_{h}|=2.88\pm 0.03$. We set the laser energy between the
two transition energies in order to excite both transitions equally
(Figure \ref{Fig3} c).

Here we observe RI with the same decay time as at zero field
(illustrated as a grey line in Fig. \ref{Fig3}b), but with a beating
pattern in the interference fringes (Figure \ref{Fig3}b). The beat
frequency corresponds to the energy difference between the allowed
transitions. These results are explained by considering two
independent 2-level systems and can be represented as a Bloch sphere
with two Bloch vectors. The two vectors precess about the z-axis at
different rates and move in and out of phase with each other. When
they are in phase we see a peak in visibility and when they are out
of phase we see a minimum. The beat visibility does not decay faster
than the exponentially decaying envelope, suggesting that the two
transitions retain mutual coherence on a timescale longer than
several nanoseconds.

\begin{figure}[H]
\includegraphics[width=160mm]{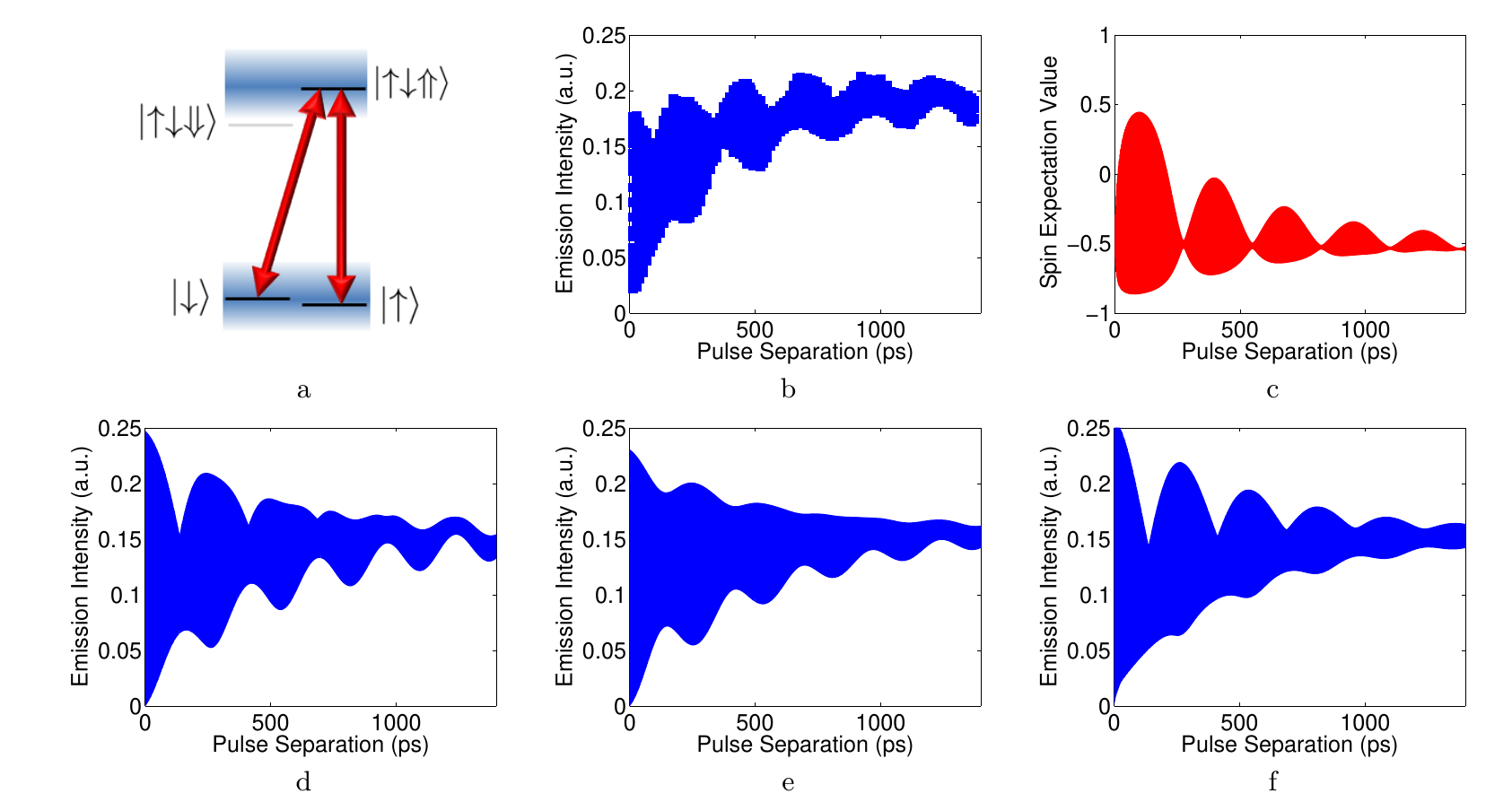}
\caption{\label{Fig5} {\bf a)} Setting the energy detuning of the
excitation laser to excite a single vertical transition results in
3-level system behaviour. {\bf b)} The measured emission intensity
of the three level system as a function of pulse separation. (The
graph appears to be a solid block of colour due to the fast
oscillations in ${\rm X}^{-}$ intensity with pulse separation. The
same is true of the simulated graphs in panels c-f.) {\bf c)} The
expectation value of the ground state electron spin in the long time
limit as a function of pulse separation. {\bf d)} A full simulation
of the expected emission intensity as a function of pulse
separation. {\bf e)} The same simulation as panel d), but here the
effect of the pulse separation-dependent spin preparation is
neglected. Note that this results in a lower beat visibility than
the experimental data, indicating that the prepared spin state plays
an important role in the expected emission intensity. {\bf f)} The
same simulation as panel d) but here the coherence terms between the
two electron ground states are rapidly destroyed. This removes the
slow oscillatory behavior at late times, indicating that the slow
oscillation is a consequence of the preparation of a coherent
superposition of the electron ground states. }
\end{figure}

Setting the laser wavelength degenerate with one of the observed
transitions in Fig. \ref{Fig3}c, we repeat the RI measurement. We
again observe a slow beating pattern with high visibility. This
time, the beat frequency corresponds to the energy of the electron Zeeman splitting. A field-dependent measurement allows us to determine the electron $g$-factor
($|g_{e}|=0.58\pm 0.08$) - in good agreement with
$|g_{e}|=0.60\pm 0.05$ made using the standard spectral method
\cite{tartakovskii2007nuclear}.

It is surprising that the ground state splitting can cause high
visibility beats in the interferogram (Figure \ref{Fig5}b). The ground state splitting is rather small so
the laser spectrum overlaps with both transitions, as shown in Fig.
\ref{Fig5}a. Although, in an ideal dot the diagonal transition in
Figure \ref{Fig5}a is forbidden, heavy-light hole mixing results in a weakly allowed
diagonal transition \cite{delteil2014observation,PhysRevB.70.241305,:/content/aip/journal/apl/97/5/10.1063/1.3473824}. However, we would
expect only a low visibility beat if this effect were resulting from
a similar mechanism to what is seen in Fig. \ref{Fig3}, namely the
independent oscillation of populations shared between vertical and
diagonal transitions, because of the difference in transition
strengths.

This high beat visibility can be explained by the dependence of the
state of the electron on the pulse separation. The system is more
likely to be excited when the electron is in the $\ket{\uparrow}$
state due to the stronger coupling of the vertical transition to
laser field. Therefore, in order to model the 3-level system we must
calculate the expected initial state of the electron.

We find that the expected initial state of the electron is
dependent on the pulse separation (Figure \ref{Fig5}c). This
dependence can be intuitively understood by
considering weak pulses using a model with two Bloch vectors (note
that this explanation neglects any coupling between the two low
energy states via the excited state). After the first pulse, the
Bloch vectors for the two transitions precess about the z-axis at
different rates. When they are completely out of phase the second
pulse can then destructively interfere with one transition and
constructively interfere with the other. The result is as if we were
preferentially driving the constructively interfering transition and
only weakly driving the destructively interfering one. After many
2-pulse repetitions, we see spin state preparation as one spin state
is being preferentially driven into the excited state. This
explains why the nodes in fringe visibility (Figure \ref{Fig5}d)
correspond to anti-nodes in the spin-expectation value graph (Figure
\ref{Fig5}c). Including the spin preparation effect in the
simulations allows us to recover the surprisingly high beat
visibility, (Figure \ref{Fig5}d). In contrast, Figure (\ref{Fig5}e)
is the result of ignoring the spin preparation - note the
considerably lower beat visibility.

As would be expected, at longer pulse separations this interference
has less of an effect and the expected spin-state becomes closer to
the calculated value for single pulses.

Next we turn our attention to the slow oscillation in the average
emission rate visible in the interferogram, a feature which is
particularly noticeable at 600-1400 ps (Figure \ref{Fig5}b). This
behavior is not seen in the 2 or 4-level systems. The oscillation
frequency corresponds to the Larmor frequency of the electron. We
determine that not only is the expected population of the electron
spin state important for determining the emission intensity, the
coherence matters too. The first pulse creates an impure coherent
superposition of the electron spin state by coupling the two spin
states via the excited state. Our simulations predict the observed
slow oscillation and show that the phase of the electron spin before
the second pulse affects the emission intensity. To emphasize this
point we destroy the electron coherence terms in the model and see
that this removes the slow oscillatory behaviour (Figure
\ref{Fig5}f). The model predicts that the visibility of these
oscillations will be higher at higher pulse powers and this is
observed in our measurements (not shown here). The simulations also
show that the decay of these oscillations at long times is only
dependent on the decay and dephasing times of the electron spin.


Our study of RI in the 3-level system demonstrates that we can
prepare a spin by controlling the phase difference between
successive pulses. This shows that it is possible to use a single
pulsed laser for both spin state preparation and exciton creation,
in contrast to work such as \cite{press2008complete} which makes use
of multiple lasers to address the spin and exciton qubits
separately. Measurements of the slow oscillation visible in Fig.
\ref{Fig5}b at longer delay times should allow direct measurement of
the electron spin coherence time. The results also show that it is
possible to manipulate the state of the system by
coherently driving the forbidden diagonal transitions of a QD in the
Faraday geometry with a resonant laser pulse. Finally, we note that
the ability to prepare and coherently manipulate an electron spin
state, along with the ability to collect
output light of all frequencies are key capabilities required to
generate electron spin-photon frequency entanglement
\cite{pmid:23151586}.

\begin{acknowledgments}
The authors acknowledge funding from the EPSRC for MBE system used in the production of the QD-LED.
J. L. gratefully acknowledges financial support from the EPSRC CDT in Photonic Systems Development and Toshiba Research Europe Ltd.
\end{acknowledgments}


%

\end{document}